\documentclass[final]{IEEEtran}
\usepackage{amsmath, amssymb}
\usepackage{etoolbox}
\usepackage{graphicx}
\usepackage{subfigure}
\usepackage[noadjust]{cite}
\newtheorem{theorem}{Theorem}
\usepackage{color}

\title{Geometry-based Estimation of Stability Region for A Class of Structure Preserving Power Grids}

\author{Thanh Long Vu and~Konstantin~Turitsyn,~\IEEEmembership{Member,~IEEE}
\thanks{Thanh Long Vu and Konstantin Turitsyn are with the Department of Mechanical Engineering, Massachusetts Institute of Technology, Cambridge, MA, 02139 USA e-mail: longvu@mit.edu and turitsyn@mit.edu.

}}

\begin{document}

\maketitle
\begin{abstract}
The increasing development of the electric power grid, the largest engineered system ever, to an even more complicated and larger system requires a new generation of stability assessment methods that are computationally tractable and feasible in real-time. In this paper we first  extend the recently introduced Lyapunov Functions Family (LFF) transient
stability assessment approach, that has potential to reduce the computational cost on  large scale power grids,
to structure-preserving power grids. Then, we introduce a new geometry-based method to 
construct the stability region estimate of power systems. Our conceptual demonstration shows that this new method can certify stability of a broader set of initial conditions compared
to the minimization-based LFF method and the energy methods (closest UEP and controlling UEP methods).
\end{abstract}

\section{Introduction}

The electrical power grid is currently undergoing the architectural revolution with the increasing penetration of renewable and distributed energy sources and the presence of millions of active endpoints. Intermittent renewable and volatile loads are difficult to exactly predict and present challenges concerning voltage, frequency, power quality, and power supply during unfavorable weather conditions. As such, the existing planning and operation computational techniques largely developed several decades ago will have to be reassessed and adopted to the new physical models
in order to ensure secure and stable operation of the modern power grids. Among those challenges, 
the extremely large size of the grid calls for the development of new generation of stability assessment methods that are computationally tractable and feasible in real-time. 

The most straightforward approach to the post-fault
stability assessment problem is based on direct time-domain
simulations of transient dynamics following the contingencies.
Rapid advances in computational hardware made it possible to
perform accurate simulations of large scale systems faster than
real-time \cite{Huang:2012il,Nagel:2013kf}. Alternatively, the direct energy approaches \cite{Pai:1981dv,Chiang:1994cUEP,Chiang:2011eo}, which are accepted and adopted by industry \cite{Tong:2010},
allow fast screening of the
contingencies while providing mathematically rigorous
certificates of stability and saving more computational resources than time-domain simulations. Essentially, the closest UEP method \cite{Chiang:1994cUEP} certifies that the post-fault dynamics is stable
if the system energy at the clearing time is smaller than the minimum energy value at every unstable equilibrium points (UEP). This method is known
conservative and not scalable to large-scale power grids since the problem of searching for an exponential number of UEPs is an NP-hard problem. The controlling UEP method \cite{Zou:2003ji} certifies that the post-fault dynamics is stable
if the system energy at the clearing time is smaller than the energy function value at the controlling UEP, which is defined as the nearest point on the boundary of the actual stability region that the fault-on trajectory is approaching, i.e. nearest the fault-cleared state. This method is less conservative than the closest UEP method since the energy value at the controlling UEP is possibly larger than the energy value at the closest UEP. However, as the actual stability region is unknown, the controlling UEP can only be searched by some heuristic algorithms.

Recently, we introduced the \emph{Lyapunov Functions Family (LFF) approach} to alleviate some of these drawbacks \cite{Vu:2014}.
The principle of this approach is to provide transient stability certificates
by constructing a family of Lyapunov functions, which are generalizations of the classical energy function,
and then find the best suited function
in the family for given initial states. Basically, this method certifies that the post-fault dynamics is stable
if the fault-cleared state stays within a polytope surrounding the equilibrium point and the Lyapunov function at the fault-cleared state is smaller than the minimum value of Lyapunov function over the flow-out boundary of that polytope. 
Generally, the LFF approach can certify stability of a
broader set of initial conditions compared to the closest UEP method.
Also, the introduced optimization-based techniques for constructing stability certificates are scalable to large-scale power grids,
since they avoid identifying the exponential number of UEPs. In addition, the LFF approach is applicable to stability assessment of 
power grids with losses \cite{Vu:2014acc}, which is impossible by the standard energy method.

In this paper, we improve the LFF transient stability assessment method and make two contributions. The first contribution is the extension of
LFF method to structure-preserving power systems. The second contribution is a new geometry-based method
to construct the estimate of stability region of the desired equilibrium point, which we argue to possibly be larger than that defined by the existing methods. We observe that among all of the UEPs, there are many points that are far from the equilibrium point and thus are not necessary to be counted when we search for the controlling UEP. Therefore, we define $2|\mathcal{E}|$ points that are the minimum points of Lyapunov function over the $2|\mathcal{E}|$ flow-out boundary segments of the considered polytope. Here, $|\mathcal{E}|$ is the number of lines in the grids. These $2|\mathcal{E}|$ minimum points play the role of all possible controlling UEPs of the system. The post-fault dynamics is certified stable
if the fault-cleared state stays within the polytope and the Lyapunov function at the fault-cleared state is smaller than the Lyapunov function at the (controlling) minimum point corresponding to the polytope's subset containing the fault-cleared state. This method is less conservative than the original LFF method in \cite{Vu:2014} since the Lyapunov function at the controlling minimum point is possibly larger than the minimum value of Lyapunov function over the flow-out boundary. In comparison to the controlling UEP method we note that since the $2|\mathcal{E}|$ minimum points play the role of all possible controlling UEPs of the system, the proposed geometry-based method can certify stability for points for which the controlling UEP method cannot. Furthermore, the construction of the minimum points is mathematically rigorous and does not involve any heuristic algorithm. Also, knowledge of the fault-on trajectory is not required as in the controlling-UEP method \cite{Zou:2003ji}.     

We note that there are many works on Lyapunov function-based stability of structure preserving power systems
\cite{bergen1981structure,Hiskens:1997Lya,hill1989lyapunov}. However, the Lyapunov function in these works is usually used to prove the local stability of the
system; it is not fully exploited to construct the stability region of the system as in this paper. Instead, in these works the stability region is estimated by the energy method.


\section{Structure Preserving Power Systems}
\label{sec:structure-preserving}

In normal conditions, power grids operate at a stable equilibrium
point. Under some fault or contingency scenarios, the system moves
away from the pre-fault equilibrium point to some post-fault
conditions. After the fault is cleared, the system experiences the
transient dynamics. This work focuses on the transient post-fault
dynamics of the power grids, and aims to develop computationally
tractable certificates of transient stability of the system, i.e.
guaranteeing that the system will converge to the post-fault
equilibrium. In this paper, we address this question on a
traditional swing equation dynamic model of power systems, which is named structure-preserving model originally introduced in \cite{bergen1981structure}.
This model naturally incorporates the dynamics of rotor angle as well as response of load power output to frequency deviation. However it does not model the dynamics of voltage in the system which is the main downside of the approach. However, in comparison to the classical swing equation with constant impedance loads, the structure of power grids is preserved in this approach.

Assume that the grid has $m$ generators and $n_0$ buses in which $n_0-m$ buses have loads and no generation. It is convenient to introduce fictitious buses representing the internal generation voltages. So, in the augmented grid we have $n=n_0+m$ buses. Assume that the grid is lossless. The $m$
generators have perfect voltage control and are characterized each
by the rotor angle $\delta_k$ and its angular velocity
$\dot\delta_k$. The dynamics of generators are described by a set
of the so-called swing equations:
\begin{align}
\label{eq.swing1}
  m_k \ddot{\delta_k} + d_k \dot{\delta_k} + P_{e_k}-P_{m_k}
  =0, k=1,..,m,
\end{align}
where, $m_k$ is the dimensionless moment of inertia of the
generator, $d_k$ is the term representing primary frequency
controller action on the governor. $P_{m_k}$ is the effective
dimensionless mechanical torque acting on the rotor and $P_{e_k}$
is the effective dimensionless electrical power output of the
$k^{th}$ generator.

Let $P_{d_k}$ be the real power drawn by the load at
$k^{th}$
bus, $k=m+1\dots,n$. In general $P_{d_k}$ is a nonlinear function of voltage and frequency. For constant voltages and small
frequency variations around the operating point $P^0_{d_k}$, it is reasonable to assume that
\begin{align}
P_{d_k}=P^0_{d_k} + d_k \dot{\delta}_k, k=m+1,\dots,n,
\end{align}
where $d_k>0$ is  the constant frequency coefficient of load. When $d_k=0$ we have the constant load model.
The electrical power $P_{e_k}$
 from the
$k^{th}$ bus into network, where $k=1,...,n,$ is given by
\begin{align}
\label{eq.electricpower}
  P_{e_k}=\sum_{j \in
  \mathcal{N}_k} V_kV_jB_{kj} \sin(\delta_k
  -\delta_j).
\end{align}
Here, the value $V_k$ represents the voltage
magnitude of the $k^{th}$ bus which is
assumed to be constant. 
$B_{kj}$ are the (normalized)  susceptance between $k^{th}$ bus and $j^{th}$ bus. $\mathcal{N}_k$ is the set of neighboring
buses of the $k^{th}$ bus. Let $a_{kj}=V_kV_jB_{kj}.$
Finally, the structure-preserving model of power systems is obtained as:
\begin{align}
\label{eq.structure-preserving}
 m_k \ddot{\delta_k} + d_k \dot{\delta_k} + \sum_{j \in
  \mathcal{N}_k} a_{kj} \sin(\delta_k-\delta_j) = &P_{m_k},  \\ 
  &k=1,\dots,m,  \nonumber\\
  \label{eq.structure-preserving2}
  d_k \dot{\delta_k} + \sum_{j \in
  \mathcal{N}_k} a_{kj} \sin(\delta_k-\delta_j) = &-P^0_{d_k},  \\ 
  & k=m+1,\dots,n. \nonumber
\end{align}

The system described by equations \eqref{eq.structure-preserving}-\eqref{eq.structure-preserving2} has many stationary points with
at least one stable corresponding to the desired operating point.
Mathematically, this point, characterized by the rotor angles
$\delta^*=[\delta_1^*,...,\delta_n^*,0,...,0]^T,$ is not unique
since any shift in the rotor angles
$[\delta_1^*+c,...,\delta_n^*+c,0,...,0]^T$ is also an
equilibrium. However, it is unambiguously characterized by the
angle differences $\delta_{kj}^*=\delta_k^*-\delta_j^*$ that solve
the following system of power-flow like equations:
\begin{align}
  \label{eq.SEP}
  \sum_{j \in
  \mathcal{N}_k} a_{kj} \sin(\delta_{kj}^*) =P_{k}, k=1,\dots,n,
\end{align}
where $P_k=P_{m_k}, k=1,\dots,m,$ and $P_k=-P^0_{d_k}, k=m+1,...,n.$
Then, the set of swing equations \eqref{eq.structure-preserving}-\eqref{eq.structure-preserving2} is
equivalent with
\begin{align}
\label{eq.swing3}
  m_k \ddot{\delta_k} + d_k \dot{\delta_k} = - \sum_{j \in
  \mathcal{N}_k} &a_{kj} \big(\sin(\delta_{kj}) -\sin(\delta_{kj}^*)\big),  \\ 
  &k=1,\dots,m,  \nonumber\\
\label{eq.swing4}  
  d_k \dot{\delta_k} = - \sum_{j \in
  \mathcal{N}_k} &a_{kj} \big(\sin(\delta_{kj}) -\sin(\delta_{kj}^*)\big),   \\ 
  & k=m+1,\dots,n. \nonumber
\end{align}

Formally, we consider the following problem.

\begin{itemize}
\item [] \textbf{Transient stability assessment problem:} \emph{Determine if the post-fault scenario defined by
initial conditions $\{\delta_k(0),\dot\delta_k(0)\}_{k=1}^n$ of the system \eqref{eq.swing3}-\eqref{eq.swing4} leads to the stable
equilibrium point $\delta^*=[\delta_1^*,...,\delta_n^*,0,...,0]^T$.}
\end{itemize}

We will address this problem by estimating the stability region of the stable equilibrium point $\delta^*,$
i.e. the set of points from which the system \eqref{eq.swing3}-\eqref{eq.swing4} will converge to $\delta^*.$ If the initial state $x_0$ belongs to this estimate set,
then the corresponding post-fault scenario is determined stable.
We will use a sequence of techniques
originating from nonlinear control theory that are most naturally
applied in the state space representation of the system. Hence, we
view the multimachine power system \eqref{eq.swing3}-\eqref{eq.swing4} as a system
with the state space vector $x = [x_1,x_2,x_3]^T$
composed of the vector of generator's angle deviations from equilibrium $x_1 =
[\delta_1 - \delta_1^*,\dots, \delta_m - \delta_m^*]^T$, their
angular velocities $x_2 = [\dot\delta_1,\dots,\dot\delta_m]^T$, and vector of load's angle deviation from equilibrium
$x_3=[\delta_{m+1}-\delta_{m+1}^*,\dots,\delta_n-\delta_n^*]^T$. 
Let $E$ be the
incidence matrix of the corresponding graph, so that
$E[\delta_1\dots\delta_n]^T =
[(\delta_k-\delta_j)_{\{k,j\}\in\mathcal{E}}]^T$. Consider matrix $C$ such that $Cx=E[\delta_1\dots\delta_n]^T.$  Consider the nonlinear transformation $F$ in this
representation is a simple trigonometric function $
F(Cx)=[(\sin\delta_{kj}-\sin\delta^*_{kj})_{\{k,j\}\in\mathcal{E}}]^T.$

In
state space representation the system can be expressed in the
following compact form:
\begin{align}
\dot{x}_1 &=x_2 \nonumber \\
\dot{x}_2 &=M_1^{-1}(-D_1x_2-S_1E^TSF(Cx))  \\
\dot{x}_3 &= -D_2^{-1}S_2E^TS F(Cx) \nonumber
\end{align}
where $S=\emph{\emph{diag}}(a_{kj})_{\{k,j\}\in \mathcal{E}}, S_1=[I_{m\times m}\quad O_{m\times n-m}], S_2=[I_{n-m\times n-m} \quad O_{n-m\times m}].$
Equivalently,
\begin{equation}\label{eq.Bilinear}
 \dot x = A x - B F(C x),
\end{equation}
with the matrices $A,B$ given by the following expression:
\begin{align}
A=\left[
        \begin{array}{ccccc}
          O_{m \times m} \qquad & I_{m \times m}  \qquad & O_{m \times n-m}\\
          O_{m \times m} \qquad & -M_1^{-1}D_1 \qquad & O_{m \times n-m} \\
          O_{m \times m} \qquad &O_{m \times m} \qquad & O_{m \times n-m}
        \end{array}
      \right],
\end{align}
and $
 B= \left[
        \begin{array}{ccccc}
          O_{m \times |\mathcal{E}|} \quad
          -M_1^{-1}S_1E^TS \quad
          -D_2^{-1}S_2E^TS
        \end{array}
      \right]^T.$
Here, $|\mathcal{E}|$ is the number of edges in the graph defined
by the susceptance matrix, or equivalently the
number of non-zero non-diagonal entries in $B_{kj}$.

\section{Lyapunov Functions Family Approach}
\label{sec:family}
\begin{figure}[t!]
\centering
\includegraphics[width = 3.2in]{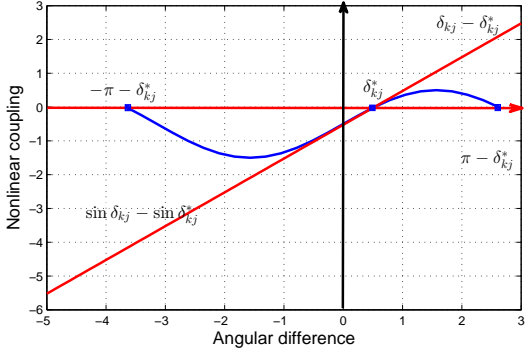}
\caption{Bounding of nonlinear sinusoidal coupling $(\sin\delta_{kj}-\sin\delta_{kj}^*)$ by two
linear functions of angular difference $\delta_{kj}$ as described in \eqref{eq.NonlinearityBouding}}
\label{fig.NonlinearityBounding}
\end{figure}
This paper proposes a family of Lyapunov functions to certify the
transient stability for the structure preserving power system \eqref{eq.Bilinear}. The
construction of this Lyapunov functions family is based on the
linear bounds of the nonlinear couplings which are clearly
separated in the state space representation \eqref{eq.Bilinear}.
From Fig. \ref{fig.NonlinearityBounding}, we observe that
\begin{align}
\label{eq.NonlinearityBouding}
  0\le
  (\delta_{kj}-\delta_{kj}^*)(\sin\delta_{kj}-\sin\delta_{kj}^*)
 \le (\delta_{kj}-\delta_{kj}^*)^2,
\end{align}
for any $| \delta_{kj}+\delta_{kj}^* |\le
\pi.$ Therefore, the nonlinearity $F(Cx)$ can be bounded by the linear functions in the polytope $\mathcal{P}$ defined by the
set of inequalities $|\delta_{kj} + \delta_{kj}^*| \le \pi$.

Exploiting this nonlinearity bounding, we propose to use the
convex cone of Lyapunov functions defined by the following system
of Linear Matrix Inequalities for positive, diagonal matrices
$K,H$ of size $2|\mathcal{E}| \times 2|\mathcal{E}|$ and
symmetric, positive matrix $Q$ of size $ 2n \times 2n:$
\begin{align}
\label{eq.QKH}
    \left[   \begin{array}{ccccc}
          A^TQ+QA  & R \\
           R^T  & -2H\\
        \end{array}\right] \le 0,
  \end{align}
where $R=QB-C^TH-(KCA)^T.$ For every pair $Q,K$ satisfying these
inequalities the corresponding Lyapunov function is given by
\begin{align}
\label{eq.Lyapunov}
V(x) = \frac{1}{2}x^T Q x &- \sum K_{\{k,j\}} (\cos\delta_{kj}+\delta_{kj}\sin\delta_{kj}^*).
\end{align}
Here, the summation goes over all elements of pair set
$\mathcal{E}$, and $K_{\{k,j\}}$ denotes the diagonal element of
matrix $K$ corresponding to the pair $\{k,j\}$.

Similar to Appendix A in \cite{Vu:2014}, we obtain the derivative of Lyapunov function $V(x)$ along
\eqref{eq.Bilinear} as:
\begin{align}
\label{eq.dotV} &\dot{V}(x)=-0.5(Xx-YF)^T(Xx-YF)
 - (Cx-F)^THF \nonumber \\
 & =-0.5(Xx-YF)^T(Xx-YF)
    - \sum H_{\{k,j\}}g_{\{k,j\}},
  \end{align}
  where $g_{\{k,j\}}=\big(\delta_{kj}-\delta_{kj}^*-(\sin\delta_{kj}-\sin\delta_{kj}^*)\big)(\sin\delta_{kj}-\sin\delta_{kj}^*).$
  From Fig. \ref{fig.NonlinearityBounding}, we have $g_{\{k,j\}} \ge 0$
  for any $|\delta_{kj}+\delta_{kj}^*|\le \pi.$ Hence, $\dot{V}(x) \le 0, \forall x \in \mathcal{P},$
  and thus the Lyapunov function is decaying in  $\mathcal{P}.$
   Therefore, we have the following result.
\begin{theorem}
\label{thr.LyapunovDecrease} \emph{In the polytope $\mathcal{P},$
the Lyapunov function defined by \eqref{eq.Lyapunov} is decaying
along the trajectory of \eqref{eq.Bilinear}, i.e., $V(x(t))$ is
decaying whenever $x(t)$ evolves inside $\mathcal{P}$.}
\end{theorem}

\section{Geometry-based Stability Certification and Contingency Screening}
\label{sec:construction}

\subsection{Construction of Stability Certificate}
In \cite{Vu:2014}, the stability certificate is constructed by finding the
minimum value $V_{\min}$ of the function $V(x)$ over the union
of flow-out boundary segments of the polytope $\mathcal{P}.$ Accordingly, if the 
Lyapunov function at the initial state, which stays inside $\mathcal{P},$ is smaller than $V_{\min},$
then the system trajectory is guaranteed to converge from the initial state to the desired stable equilibrium point.
In this paper, we will introduce a geometry-based approach for stability certificate construction,
in which we inscribe  inside the polytope $\mathcal{P}$ an invariant set $\mathcal{R}$ which is the largest set formed by combining the flow-in boundary of the polytope $\mathcal{P}$ together  with the patches of Lyapunov function's sublevel sets that are guaranteed do not meet the flow-out boundary of $\mathcal{P}$.

In deed, we divide the
boundary $\partial\mathcal{P}_{kj}$ of $\mathcal{P}$ corresponding to the
equality $|\delta_{kj}+\delta_{kj}^*|=\pi$ into two
subsets $\partial\mathcal{P}_{kj}^{in}$ and
$\partial\mathcal{P}_{kj}^{out}$. The flow-in boundary
segment $\partial\mathcal{P}_{kj}^{in}$ is defined by
$|\delta_{kj}+\delta_{kj}^*|=\pi$ and
$\delta_{kj}\dot{\delta}_{kj} < 0,$ while the flow-out boundary
segment $\partial\mathcal{P}_{kj}^{out}$ is defined by
$|\delta_{kj}+\delta_{kj}^*|=\pi$ and
$\delta_{kj}\dot{\delta}_{kj} \ge 0.$ Since the derivative of $\delta_{kj}^2$ at every points on
$\partial\mathcal{P}_{kj}^{in}$ is negative, the system trajectory can only go inside $\mathcal{P}$ once it meets $\partial\mathcal{P}_{kj}^{in}.$
We define the following minimum values of $V(x)$ on the flow-out boundary segment
$\partial\mathcal{P}_{kj}^{out}$:
\begin{align}\label{eq.Vmin1}
 V^{\pm}_{\min_{kj}}=\mathop {\min}\limits_{x \in \partial\mathcal{P}_{kj}^{out\pm}} V(x),
\end{align}
where $\partial\mathcal{P}_{kj}^{out\pm}$ is the flow-out boundary
segment of polytope $\mathcal{P}$ that is defined by $\delta_{kj}
+\delta_{kj}^* = \pm\pi$ and $\delta_{kj}\dot{\delta}_{kj} \ge 0$.
Let $x_{kj}^{out\pm}$ be the point on $\partial\mathcal{P}_{kj}^{out\pm}$ such that
$V(x_{kj}^{out\pm})=V^{\pm}_{\min_{kj}}.$ 

\begin{figure}
\centering
\includegraphics[width = 3.2in]{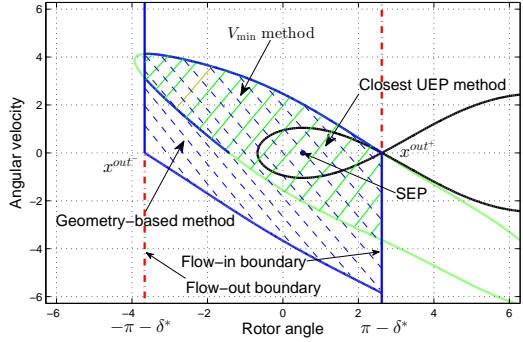}
\caption{Comparison between the stability region estimates defined by $V_{\min}$-based method and
geometry-based method with the stability region  obtained
by the closest UEP energy method (black solid line). The stability region estimated by the $V_{\min}$ method is the intersection of the Lyapunov level set (blue 
solid line) and the polytope defined by $-\pi-\delta^* \le \delta
\le \pi-\delta^*.$ The stability region estimated by the geometry-based method is the inner of the set whose boundary is combined of solid blue segments} \label{fig.comparison_2Bus}
\end{figure}

Now we consider the set $\mathcal{R}$ formed by the combination of the flow-in boundary $\partial\mathcal{P}^{in}$ of the polytope $\mathcal{P}$
together with $2|\mathcal{E}|$ segments of Lyapunov function's sublevel sets. Each of these segments goes through one of the $2|\mathcal{E}|$  points $x_{kj}^{out\pm}$
and lies in the half of the polytope $\mathcal{P}$ corresponding to $\emph{\emph{sign}}(\dot{\delta}_{kj})=\pm.$ The conceptual demonstration of the set $\mathcal{R}$ is given as
the combination of solid blue lines
in Fig. \ref{fig.comparison_2Bus}. Note, these segments 
can only meet the boundary of $\mathcal{P}$ at the point with $\delta_{kj}\dot{\delta}_{kj}=(\mp \pi- \delta_{kj}^*)(\pm)<0,$ i.e. the point on the flow-in boundary. Therefore, the boundary of $\mathcal{R}$ is composed of segments which are parts of Lyapunov function's sublevel sets or flow-in boundary. 

From the decrease of Lyapunov function inside $\mathcal{P}$ (Theorem \ref{thr.LyapunovDecrease}) we note that from any initial state inside $\mathcal{R}$ the system trajectory cannot escape $\mathcal{R}$ through the Lyapunov function's sublevel sets. Also, once the system trajectory meets the flow-in boundary, it can only go inside the polytope $\mathcal{P}.$ So, if the set $\mathcal{R}$ is closed, then its inner is an invariant set.   
In Appendix \ref{appen:estimate}, we prove the following main result of this paper.

\begin{theorem}
\label{thr.stabilityregion}
\emph{If the set $\mathcal{R}$ is closed \footnote{We conjecture that there are always some Lyapunov functions in the family defined by the LMIs \eqref{eq.QKH} such that the set
$\mathcal{R}$ is closed. In the conceptual demonstration of 2-bus system, it is easy to search for such Lyapunov function by the adaptation algorithm introduced in \cite{Vu:2014}.}, then the inner of $\mathcal{R}$ is an estimate of the stability region of the equilibrium point $\delta^*,$ i.e., 
from any initial state $x_0$ in the set $\mathcal{R},$ the system trajectory $x_t$ of \eqref{eq.Bilinear} will converge to  
$\delta^*.$}
\end{theorem}

Theorem \ref{thr.stabilityregion} provides a geometry-based estimate of the stability region of the stable equilibrium point. As a conceptual illustration, we can observe from Fig. \ref{fig.comparison_2Bus} that in the most simple case of 2-bus system, the geometry-based method results in the largest stability region estimate compared to the closest UEP method and the $V_{\min}$ method in \cite{Vu:2014}.

\subsection{Direct Method for Contingency Screening}

\label{sec.screen} In this section, we will apply the geometry-based stability certificate to the contingency screening problem. Essentially, the post-fault dynamics is certified stable
if the fault-cleared state $x_0$ stays within the polytope $\mathcal{P}$ and the Lyapunov function at $x_0$ is smaller than the Lyapunov function at the (controlling) minimum point corresponding to the polytope's subset containing the fault-cleared state. Indeed, for a given fault-cleared state $x_0,$
which is determined by integration or other techniques, the value of $V(x_0)$ can be computed by direct application of
\eqref{eq.Lyapunov}. If $x_0$ is inside the polytope $\mathcal{P},$ we calculate the frequency differences $\dot{\delta}_{kj}.$ From the $|\mathcal{E}|$ signatures of these frequency differences,
we can determine the subset of the polytope $\mathcal{P}$ in which every points have the same signatures for frequency differences with $x_0.$ Then, from the formulation \eqref{eq.Vmin1} we can define $|\mathcal{E}|$
minimum values $V^{\pm}_{\min_{kj}}$, in which $V^{\pm}_{\min_{kj}}$ is either $V^{+}_{\min_{kj}}$ or $V^{-}_{\min_{kj}}$ according to the signature of $\dot{\delta}_{kj}.$ 
The value of Lyapunov function at the initial state $x_0$ should be then compared to the minimum
of these $|\mathcal{E}|$
minimum values $V^{\pm}_{\min_{kj}}$. If $V_0$ is smaller than this minimum value, the post-fault dynamics is certified stable, because $x_0$ belongs to the stability region estimate $\Phi$.

We note that unlike energy based approaches, the LFF method provides a whole
cone of Lyapunov functions to choose from. This freedom can be
exploited to choose the Lyapunov function that is best suited for
a given initial condition or their family. Essentially, we can apply the similar
 iterative algorithm in \cite{Vu:2014} (Section IV) to identify the Lyapunov
function that certifies the stability of a given initial condition
$x_0$ whenever such a Lyapunov function exits.

\section{Simulation Results}
\label{sec:simulation}

To illustrate the effectiveness of the LFF and geometry-based approach in estimating the stability region of power systems, we consider the classical $2$-bus with easily visualizable
state-space regions. This system is described by a single 2-nd
order differential equation
\begin{align}
  m \ddot{\delta} +d \dot{\delta} + a \sin\delta - P=0.
\end{align}
For this system $\delta^*=\arcsin(P/a)$ is the only stable
equilibrium point (SEP). For numerical simulations, we choose
$m=1$ p.u., $d=1$ p.u., $a= 0.8$ p.u., $P=0.4$ p.u., and
$\delta^*=\pi/6.$ Figure \ref{fig.comparison_2Bus} illustrates the construction of stability region estimate for the most simple 2-bus system by the closest UEP method, the $V_{\min}$ method in \cite{Vu:2014},
and the geometry-based method. It can be seen that there are many
contingency scenarios defined by the configuration $x_0$ whose
stability cannot be certified by the energy method, but
can be ensured by the LFF method. Also, the geometry-based method provides a better stability region estimate 
compared to the $V_{\min}$ method.

We can also see that the two minimum points $x^{out\pm}$ are all the UEPs of the system. Hence, the estimate set $\mathcal{R}$ covers the Lyapunov function's sublevel sets that go through the UEPs. Therefore, the geometry-based stability certificate can assess transient stability for every initial states in $\mathcal{P}$ that the controlling UEP method in \cite{Zou:2003ji} does.

\section{Conclusions and Path Forwards}
\label{sec:discussion}
This paper extended the recently introduced LFF approach to transient stability certification of structure-preserving power systems.
A new geometry-based technique was also introduced to further enlarge the estimate of stability region compared to the original LFF method. The new estimate is the largest set formed by combining the flow-in boundary of the polytope in which the Lyapunov function is decreasing together  with the patches of sublevel sets that are guaranteed do not meet the flow-out boundary of that polytope.
Our numerical simulations showed that this new estimate of stability region is broader than that obtained by
 the energy methods and the original LFF method. In the applications to contingency screening,
the geometry-based technique in this paper resulted in a more complicated algorithm compared to the original LFF method in \cite{Vu:2014}.
However, the larger stability region estimate obtained by the geometry-based method guaranteed that more contingency scenarios are screened and certified stable.

Toward the practical applications of the Lyapunov Functions Family approach to transient stability certification, further extensions should be made in the future where more complicated structure-preserving models of power systems are considered,
  e.g. the dynamics of generators' voltage or effects of buses' reactive power is incorporated in the model.
Since the LFF method is applicable to lossy power grid \cite{Vu:2014acc}, it is straightforward to extend the method to incorporating reactive power,
which will introduce the cosine term in the model \eqref{eq.swing3}. This can be done by extending the state vector $x$ and combining the technique in this paper with the LFF transient stability
techniques in \cite{Vu:2014acc} for lossy power grids (without reactive power considered). Also, we can see from the proof of Theorem \ref{thr.LyapunovDecrease}
that, in order to make sure the Lyapunov function is decreasing in the polytope $\mathcal{P},$ it is not necessary to restrict the nonlinear terms $F(Cx)$
to be univariate. As such, we can extend the LFF method to power systems with generators' voltage dynamics in which the voltage variable is incorporated in a multivariable nonlinear function $F.$

We envision to develop a new security assessment toolbox for practical power grids based on the LFF approach. This tool can certify transient stability for a broad set of contingency scenarios when the dynamics of power systems in described by a number of models, from simple classical reduction model to complex
structure-preserving model with dynamic voltage and reactive power incorporated. Also, this security assessment toolbox can certify stability for rather complicated situations when the system parameters are changing or unknown via the robust stability certificate developed in \cite{Vu:2014acc}. We will build a library of models and contingency scenarios the stability of which can be certified by this security assessment toolbox. This will help us quickly assess the transient stability of dynamical power systems by offline algorithms.  

\section{Acknowledgements}
This work was partially supported by MIT/Skoltech and
Masdar initiatives.

\section{Appendix}
\subsection{Proof of Theorem \ref{thr.stabilityregion} for Stability Region Estimate}
\label{appen:estimate}

Since inner of $\mathcal{R}$ is an invariant set we have $x(t) \in \mathcal{R} \subset \mathcal{P}$ for all $t\ge 0.$ By Theorem \ref{thr.LyapunovDecrease}
we have $\dot{V}(x(t)) \le 0$ for all $t.$ From LaSalle theorem, we conclude that the system trajectory $x(t)$ will converge to the 
set $\{x:\dot{V}(x)=0\}.$ This together with \eqref{eq.dotV} imply that the system trajectory will converge to the stable equilibrium point $\delta^*$
or to some point lying on the boundary of $\mathcal{P}.$ However, by the construction of $\mathcal{R}$ the second case cannot happen. Therefore, the system 
will converge to $\delta^*.$

\bibliographystyle{IEEEtran}
\bibliography{lff}

\newcommand{\noopsort}[1]{} \newcommand{\printfirst}[2]{#1}
  \newcommand{\singleletter}[1]{#1} \newcommand{\switchargs}[2]{#2#1}
\begin{thebibliography}{10}
\providecommand{\url}[1]{#1}
\csname url@samestyle\endcsname
\providecommand{\newblock}{\relax}
\providecommand{\bibinfo}[2]{#2}
\providecommand{\BIBentrySTDinterwordspacing}{\spaceskip=0pt\relax}
\providecommand{\BIBentryALTinterwordstretchfactor}{4}
\providecommand{\BIBentryALTinterwordspacing}{\spaceskip=\fontdimen2\font plus
\BIBentryALTinterwordstretchfactor\fontdimen3\font minus
  \fontdimen4\font\relax}
\providecommand{\BIBforeignlanguage}[2]{{%
\expandafter\ifx\csname l@#1\endcsname\relax
\typeout{** WARNING: IEEEtran.bst: No hyphenation pattern has been}%
\typeout{** loaded for the language `#1'. Using the pattern for}%
\typeout{** the default language instead.}%
\else
\language=\csname l@#1\endcsname
\fi
#2}}
\providecommand{\BIBdecl}{\relax}
\BIBdecl

\bibitem{Huang:2012il}
Z.~Huang, S.~Jin, and R.~Diao, ``{Predictive Dynamic Simulation for Large-Scale
  Power Systems through High-Performance Computing},'' \emph{High Performance
  Computing, Networking, Storage and Analysis (SCC), 2012 SC Companion}, pp.
  347--354, 2012.

\bibitem{Nagel:2013kf}
I.~Nagel, L.~Fabre, M.~Pastre, F.~Krummenacher, R.~Cherkaoui, and M.~Kayal,
  ``{High-Speed Power System Transient Stability Simulation Using Highly
  Dedicated Hardware},'' \emph{Power Systems, IEEE Transactions on}, vol.~28,
  no.~4, pp. 4218--4227, 2013.

\bibitem{Pai:1981dv}
M.~A. Pai, K.~R. Padiyar, and C.~RadhaKrishna, ``{Transient Stability Analysis
  of Multi-Machine AC/DC Power Systems via Energy-Function Method},''
  \emph{Power Engineering Review, IEEE}, no.~12, pp. 49--50, 1981.

\bibitem{Chiang:1994cUEP}
H.-D. Chiang, F.~F. Wu, and P.~P. Varaiya, ``{A BCU method for direct analysis
  of power system transient stability },'' \emph{Power Systems, IEEE
  Transactions on}, vol.~9, no.~3, pp. 1194--1208, 1994.

\bibitem{Chiang:2011eo}
H.-D. Chiang, \emph{{Direct Methods for Stability Analysis of Electric Power
  Systems}}, ser. Theoretical Foundation, BCU Methodologies, and
  Applications.\hskip 1em plus 0.5em minus 0.4em\relax Hoboken, NJ, USA: John
  Wiley {\&} Sons, Mar. 2011.

\bibitem{Tong:2010}
J.~Tong, H.-D. Chiang, and Y.~Tada, ``{On-line power system stability screening
  of practical power system models using TEPCO-BCU},'' in \emph{ISCAS}, 2010,
  pp. 537--540.

\bibitem{Zou:2003ji}
Y.~Zou, M.-H. Yin, and H.-D. Chiang, ``{Theoretical foundation of the
  controlling UEP method for direct transient-stability analysis of
  network-preserving power system models},'' \emph{Circuits and Systems I:
  Fundamental Theory and Applications, IEEE Transactions on}, vol.~50, no.~10,
  pp. 1324--1336, 2003.

\bibitem{Vu:2014}
T.~L. Vu and K.~Turitsyn, ``{Lyapunov functions family approach to transient
  stability assessment},'' \emph{Power Systems, IEEE Trans.}, 2014, in review,
  available: arXiv:1409.1889.

\bibitem{Vu:2014acc}
------, ``{Synchronization stability of lossy and uncertain power grids},'' in
  \emph{2015 American Control Conference}, accepted.

\bibitem{bergen1981structure}
A.~R. Bergen and D.~J. Hill, ``A structure preserving model for power system
  stability analysis,'' \emph{Power Apparatus and Systems, IEEE Transactions
  on}, no.~1, pp. 25--35, 1981.

\bibitem{Hiskens:1997Lya}
R.~Davy and I.~A. Hiskens, ``{Lyapunov functions for multi-machine power
  systems with dynamic loads},'' \emph{Circuits and Systems I: Fundamental
  Theory and Applications, IEEE Transactions on}, vol.~44, 1997.

\bibitem{hill1989lyapunov}
D.~J. Hill and C.~N. Chong, ``Lyapunov functions of lur'e-postnikov form for
  structure preserving models of power systems,'' \emph{Automatica}, vol.~25,
  no.~3, pp. 453--460, 1989.

\end{thebibliography}
\end{document}